\documentclass[11pt, letterpaper]{article}
\pdfoutput=1

\usepackage[utf8]{inputenc}
\usepackage[T1]{fontenc}
\usepackage[margin=1in]{geometry}
\usepackage{hyperref}
\usepackage{url}
\usepackage{booktabs}
\usepackage{xcolor}
\usepackage{microtype}
\usepackage{fancyhdr}
\usepackage{titlesec}
\usepackage{listings}
\usepackage{tcolorbox}
\usepackage{array}
\usepackage{tabularx}
\usepackage{multirow}
\usepackage{natbib}
\usepackage{graphicx}
\usepackage{amsmath}
\usepackage{amssymb}
\usepackage{setspace}
\usepackage{parskip}
\usepackage{pgfplots}
\pgfplotsset{compat=1.18}

\definecolor{aira-blue}{RGB}{31,78,121}
\definecolor{aira-mid}{RGB}{46,117,182}
\definecolor{aira-light}{RGB}{232,240,248}
\definecolor{aira-warn}{RGB}{255,248,232}
\definecolor{aira-code}{RGB}{244,244,244}
\definecolor{aira-red}{RGB}{192,0,0}

\hypersetup{
  colorlinks=true,
  linkcolor=aira-mid,
  citecolor=aira-mid,
  urlcolor=aira-mid,
  pdftitle={AIRA -- AI-Induced Risk Audit},
  pdfauthor={William M. Parris},
  pdfsubject={Software Engineering, AI Safety},
  pdfkeywords={AI-generated code, fail-soft, RLHF, static analysis, software reliability}
}

\titleformat{\section}{\large\bfseries\color{aira-blue}}{{\thesection}}{1em}{}
\titleformat{\subsection}{\normalsize\bfseries\color{aira-mid}}{{\thesubsection}}{1em}{}
\titleformat{\subsubsection}{\normalsize\bfseries\itshape\color{aira-mid}}{{\thesubsubsection}}{1em}{}

\tcbuselibrary{skins,breakable}

\newenvironment{quotebox}{%
  \begin{tcolorbox}[
    enhanced,
    colback=aira-light,
    colframe=aira-mid,
    leftrule=4pt, rightrule=0pt, toprule=0pt, bottomrule=0pt,
    arc=0pt,
    left=6pt, right=6pt, top=4pt, bottom=4pt
  ]
  \itshape
}{%
  \end{tcolorbox}
}

\newenvironment{definitionbox}{%
  \begin{tcolorbox}[
    enhanced,
    colback=aira-light!60,
    colframe=aira-blue,
    arc=2pt,
    left=8pt, right=8pt, top=6pt, bottom=6pt
  ]
}{%
  \end{tcolorbox}
}

\newcommand{\checkid}[1]{\texttt{#1}}
\newcommand{\aira}{\textsc{aira}}
\newcommand{\HIGH}{\textcolor{aira-red}{\textbf{HIGH}}}
\newcommand{\MED}{\textcolor{orange}{\textbf{MEDIUM}}}
\newcommand{\LOW}{\textcolor{gray}{\textbf{LOW}}}

\title{\vspace{-1.5cm}\textbf{\color{aira-blue} AIRA --- AI-Induced Risk Audit} \\ \Large\itshape A Structured Inspection Framework for AI-Generated Code}
\author{\textbf{William M.~Parris} \\ \normalsize BDB Labs $\mid$ Jurisprudential AI Governance Initiative \\ \normalsize \href{mailto:bill@bageltech.net}{bill@bageltech.net} $\quad$ \href{https://aira.bageltech.net}{aira.bageltech.net}}
\date{\normalsize 2026}

\begin{document}

\maketitle

% ── Abstract ──────────────────────────────────────────────────────────────────
\begin{abstract}
Practitioners have reported a directional pattern in AI-assisted code generation:
AI-generated code tends to fail quietly, preserving the surface appearance of function
while degrading or concealing guarantees. This paper introduces the
\emph{Reward-Shaped Failure Hypothesis} --- the proposal that this pattern may reflect an artifact
of how AI systems are optimized through human feedback, rather than a random distribution
of bugs. Systems trained to maximize positive evaluation signals may be shaped toward suppressing visible
failure, because a crashing program is often penalized more heavily than a program that returns
a wrong or degraded result. We define \emph{failure truthfulness} as the property that a
system's observable outputs accurately represent its internal success or failure state.
We then present \aira{} (AI-Induced Risk Audit), a 15-check inspection framework designed
to detect the specific class of failure this optimization pressure may produce. \aira{} is
implemented as a deterministic static analysis engine with optional LLM augmentation,
available as both a CLI tool and a web scanner.

We report empirical results from three studies: an anonymized six-system enterprise
AI-assisted environment audit that operationally characterizes where the hypothesis emerged;
a rebuilt balanced 600-file public-corpus pilot; and a stricter matched-control replication
comparing 955 AI-attributed files against 955 human-control files. In the final replication,
AI-attributed files show 0.435 high-severity findings per file versus 0.242 in human
controls, a 1.80$\times$ excess, with the same direction observed in JavaScript, Python,
and TypeScript. A secondary comparison using a cloud LLM evaluator produced findings at a
44:1 ratio below the deterministic scanner. \aira{} is designed for governance,
compliance, and safety-critical systems where fail-closed behavior is a definitional
requirement.
\end{abstract}
% ─────────────────────────────────────────────────────────────────────────────
\section{Introduction}
% ─────────────────────────────────────────────────────────────────────────────

The question traditionally asked of software is whether it works.
Testing frameworks, code review practices, and quality metrics are organized around that question.
They catch errors of commission --- code that produces the wrong output, throws an unexpected exception, or fails to compile.
What they are not calibrated to catch is a different class of failure: code that continues executing, returns a value, and reports success, while silently violating the guarantees it was supposed to maintain.
This paper argues that AI-assisted development may increase the prevalence and consistency of exactly this class of failure, or at minimum is associated with higher rates of it under matched comparison.

The argument is structural, not anecdotal. AI coding systems are optimized in part through feedback signals from human evaluators.
A program that crashes produces a strongly negative signal. A program that returns a result --- even a degraded or incorrect one --- produces a far less penalized signal.
Over training, this asymmetry may shape code generation toward the \emph{appearance} of correctness rather than correctness under all conditions.
The result is not a random distribution of bugs. It is a directional skew: AI-generated code disproportionately fails in ways that preserve surface function while concealing broken guarantees.
Standard review practices, designed for human error patterns, are not calibrated to detect it.

We introduce \aira{} (AI-Induced Risk Audit) as a targeted response to this problem. \aira{} is not a general-purpose code quality tool.
It is a structured inspection framework organized around a single question:

\begin{quotebox}
Not ``does this code work?'' but ``does this code tell the truth about whether it is working?''
\end{quotebox}

The remainder of this paper is organized as follows.
Section~\ref{sec:theory} develops the Reward-Shaped Failure Hypothesis and formally defines failure truthfulness.
Section~\ref{sec:related} positions \aira{} relative to existing work.
Section~\ref{sec:framework} presents the framework design, architecture, and 15 checks.
Section~\ref{sec:empirical} reports empirical results from three studies.
Section~\ref{sec:output} describes the output specification.
Section~\ref{sec:limitations} addresses limitations.
Section~\ref{sec:conclusion} concludes with future directions.

\subsection{Contributions}

This work makes the following contributions:
\begin{itemize}
  \item Introduces the \emph{Reward-Shaped Failure Hypothesis} --- a proposed structural account of why AI-assisted codebases exhibit systematic fail-soft behavior
  \item Defines \emph{failure truthfulness} as a formal system property distinct from correctness
  \item Presents \aira{}, a 15-check inspection framework targeting the specific failure class this hypothesis predicts
  \item Reports empirical findings from a deterministic \aira{} audit of a six-system enterprise AI-assisted development environment
  \item Reports a rebuilt balanced 600-file corpus pilot (300 AI-attributed vs.\ 300 human-authored) showing a 1.32$\times$ overall and 3.18$\times$ JavaScript high-severity differential
  \item Reports a stricter matched-control replication of 955 AI-attributed files versus 955 human-control files, showing a 1.80$\times$ high-severity differential with the same direction across JavaScript, Python, and TypeScript
  \item Demonstrates that an LLM evaluator applied to the same surfaces exhibits the same suppression pattern at a 44:1 ratio below deterministic detection
  \item Releases \aira{} as an open-source tool with CLI, web, and research collection components at \url{https://github.com/BDB-Labs/aira-scanner}
\end{itemize}

% ─────────────────────────────────────────────────────────────────────────────
\section{Theoretical Foundation}
\label{sec:theory}
% ─────────────────────────────────────────────────────────────────────────────

\subsection{The Pattern}

Engineers who work extensively with AI coding tools begin to notice a pattern.
The code compiles. The tests pass. The function returns a value.
And yet, under failure conditions, the system does not stop --- it continues, quietly, in a degraded state that nothing in the output makes visible.
Errors are absorbed. Exceptions are swallowed. Guarantees written into the design simply do not propagate when they are violated.

This is a consistent, directional skew: AI-generated code fails quietly, in ways that preserve the surface appearance of function.
Standard code review practices are not calibrated to detect this class of failure, because human developers do not produce it at the same rate or in the same form.
In a coding context, this pattern most commonly manifests as:
\begin{itemize}
  \item \textbf{Silent exception handling} --- because a swallowed error looks like continued success
  \item \textbf{Graceful degradation without disclosure} --- because a degraded response feels better than a hard stop
  \item \textbf{Optimistic return values} --- because returning \texttt{None} or raising feels like giving up
  \item \textbf{Broad try/catch blankets} --- because a crash is the worst possible visible outcome
  \item \textbf{Happy-path test coverage} --- because passing tests generate positive feedback
\end{itemize}

\subsection{The Reward-Shaped Failure Hypothesis}

\aira{} offers a hypothesis about the structural cause.
AI systems are optimized, in part, through feedback signals that reward outputs humans evaluate positively.
A crashing program is a strongly negative signal. A program that returns a result --- even a wrong one, even a degraded one --- is far less penalized.
Over training, this pressure shapes how these systems generate code: not toward correctness under all conditions, but toward the \emph{appearance} of correctness under the conditions most likely to be evaluated.
In code-generation settings, those evaluation conditions often reward scripts that run to completion in a simple sandbox or satisfy a shallow success criterion;
swallowing an exception or returning a fallback can therefore score better with human raters than surfacing an explicit failure, even when the underlying guarantee has been broken.

\begin{quotebox}
The failure patterns documented in this framework are consistent with an optimization-shaped pressure in the feedback environment, rather than a defect in any particular implementation.
\end{quotebox}

Because the failure distribution is non-random, it is auditable. The same optimization pressure that appears to produce silent exception handling in one codebase may produce similar patterns in another.
\aira{} is organized around that predictability.

This hypothesis connects two conversations that have largely proceeded in parallel.
Software reliability researchers have documented increasing rates of silent failure in AI-assisted codebases.
AI safety researchers have identified that RLHF inherently trains models toward fail-open behavior \citep{casper2023open}.
\aira{} bridges these observations by proposing that the alignment process is a plausible structural contributor to the software vulnerability.

\subsection{Failure Truthfulness}
\label{sec:failure-truthfulness}

The innovation in \aira{} is not detection of bugs per se, but detection of a specific epistemic failure.
We define:

\begin{definitionbox}
  \textbf{Failure Truthfulness:} The property that a system's observable outputs accurately
  represent its internal success or failure state, without suppression, ambiguity, or
  degradation masking.
\end{definitionbox}

A system can be functionally correct on the happy path and \emph{failure-untruthful} under error conditions.
Standard correctness testing does not distinguish these. \aira{} is organized around failure truthfulness as a first-class system property, separate from and complementary to correctness.
Failure truthfulness fails when a system returns a success signal after a critical operation has failed (Check C01), when audit evidence can be lost without halting execution (C02), when exceptions are absorbed without preserving failure semantics (C03), or when outputs are presented as authoritative without a confidence posture (C13).
These are not independent bugs --- they are different surfaces of the same underlying property violation.

\subsection{Implications for Governance-Critical Systems}

The stakes are highest in governance, compliance, and safety-critical systems --- where fail-closed behavior is not a design preference but a definitional requirement.
A system that must refuse to act under uncertainty cannot be built reliably on code that was shaped to suppress uncertainty signals.
\aira{} was developed in part through direct observation of this tension during construction of production Constitutional AI governance frameworks, where AI coding agents consistently introduced silent exception handlers into audit-critical paths --- not through any individual mistake, but as a systematic pattern across hundreds of generated functions.
In distributed systems, the failure mode compounds. A swallowed exception in a microservice does not only affect a local operation --- it can corrupt downstream data pipelines, poison distributed caches, and propagate a false success signal across system boundaries.

% ─────────────────────────────────────────────────────────────────────────────
\section{Related Work and Positioning}
\label{sec:related}
% ─────────────────────────────────────────────────────────────────────────────

\aira{} occupies a specific position in the landscape of code quality and AI safety tooling.
General-purpose static analysis tools --- SonarQube, Semgrep, CodeClimate, Pylint --- detect broad classes of defect including security vulnerabilities, style violations, and known bug patterns.
They are not designed to detect the directional failure bias introduced by optimization pressure.
An AI-generated codebase can pass standard static analysis while exhibiting pervasive fail-soft behavior, because the failure mode does not primarily manifest as syntactic error or known vulnerability pattern --- it manifests as structurally valid code with the wrong semantics under failure conditions.

Observability and failure transparency research has long recognized the distinction between fail-open and fail-closed system design \citep{beyer2016sre}.
\aira{} formalizes these concerns as an auditable checklist and connects them to the hypothesized structural cause in AI-assisted development.

Research on LLM code generation quality has documented that AI-generated code exhibits higher rates of missing boundary checks, empty exception handlers, and reduced robustness compared to human-written equivalents.
Large-scale structural comparisons of agentic versus human-authored pull requests using the AIDev dataset have confirmed that AI-generated code is structurally distinct from human-written code across commit structure and breadth of file modifications \citep{ogenrwot2026how}.
Empirical analysis of test failures in AI-generated PRs has found that runtime errors dominate at 62.6\% versus 37.4\% compile-time failures, with assertion failures as the most common issue \citep{msr2026testfailures} --- a distribution consistent with fail-soft behavior.
In AI safety, the fail-open tendency of RLHF-trained models has been discussed primarily in the context of conversational evasion \citep{casper2023open}.
\aira{} applies this observation to software architecture.

\subsection{Distinction from Prior Work}

While recent studies have documented higher rates of robustness issues and bugs in LLM-generated code --- such as missing boundary checks, empty exception handlers, and reduced performance under edge cases --- most focus on measuring bug prevalence or functional correctness without identifying a unifying structural cause or providing targeted auditing instrumentation.
Benchmarks such as CoderEval \citep{yu2024codereval} have evaluated pragmatic code generation and shown varying robustness, while systematic surveys catalog functional, security, and other bug types in AI-generated code \citep{gao2025survey}.

\aira{} advances the literature by (1) naming the Reward-Shaped Failure Hypothesis as the underlying optimization-driven mechanism, (2) defining failure truthfulness as a distinct, measurable system property complementary to traditional correctness, and (3) delivering a deterministic 15-check framework explicitly engineered to detect the predicted failure class while remaining resistant to the same suppression tendencies.
This bridges software reliability and observability research with AI alignment concerns in a directly actionable way for governance-critical systems.

% ─────────────────────────────────────────────────────────────────────────────
\section{The AIRA Framework}
\label{sec:framework}
% ─────────────────────────────────────────────────────────────────────────────

\subsection{Design Principles}

\aira{} was designed around four principles that distinguish it from general code quality tools:

\textbf{Hypothesis-derived checks.} The 15 checks target specific failure modes that the Reward-Shaped Failure Hypothesis predicts.
The framework is organized around a single question: does this code tell the truth about whether it is working?

\textbf{Deterministic backbone.} The primary scanner is parser-backed deterministic analysis --- not LLM-assisted classification.
This ensures the instrument itself is not susceptible to the failure mode it is designed to detect.
LLM augmentation is available but treated as optional enrichment, not ground truth.

\textbf{Fail-closed semantics.} \texttt{UNKNOWN} is not a neutral result.
Two checks (C07, C12) require human review. On governance-critical paths, \texttt{UNKNOWN} should be treated as a conditional \texttt{FAIL} pending manual verification.

\textbf{Research posture.} \aira{} makes no claim that a passing result means a system is safe.
The correct language for \aira{} outputs is \emph{observed}, \emph{measured}, and \emph{suggests} --- not \emph{proves} or \emph{guarantees}.

\textbf{What \aira{} is not.} \aira{} is not a general-purpose static analyzer: it does not detect security vulnerabilities, style violations, or broad correctness defects.
It is not an authorship detector: it identifies failure-untruthful patterns regardless of whether AI or a human wrote the code.
It is not a hallucination detector: it does not evaluate model outputs for factual accuracy.
And it is not a claim that every flagged pattern is a defect --- some fail-soft behavior is intentional in context.
\aira{} is specifically a \emph{failure-truthfulness} instrument: it surfaces conditions under which a system may misrepresent its own correctness state.

\subsection{System Architecture}

\aira{} is implemented as four connected components:

\textbf{Deterministic rule engine.} Parser-backed static analysis for Python (AST-based), JavaScript, TypeScript, and JSX/TSX (esprima-backed with lexical fallback).

\textbf{CLI tool.} Supports static, LLM, and hybrid scan modes across local repos.
Includes Ollama model discovery, selected-model validation, and aggregate-only research submission via \texttt{--submit-research-aggregate}. Ollama is treated as a generic abstraction layer.

\textbf{Web scanner.} Provider-routed scan surface with fallback hierarchy: configured cloud or Ollama route $\to$ deterministic server-side static scan $\to$ browser heuristics.

\textbf{Research collection pipeline.} Aggregate-only submission to Supabase (hosted) or JSONL (local/CI). Raw source code, file paths, and snippets are never transmitted.

\subsection{Scan Modes}

\aira{} supports three scan modes. \emph{Static} mode uses deterministic built-in analysis and is the canonical baseline for research.
\emph{LLM} mode uses provider-assisted analysis and is useful for exploratory comparison.
\emph{Hybrid} mode merges static and LLM findings, allowing provider-assisted signal to supplement the deterministic baseline.

\subsection{The 15 Checks}
\label{sec:checks}

Each check targets a specific failure mode predicted by the Reward-Shaped Failure Hypothesis.
Table~\ref{tab:checks} summarizes the full check set.

\begin{table}[htbp]
\centering
\caption{AIRA Check Summary}
\label{tab:checks}
\small
\begin{tabular}{@{}llp{4cm}p{5cm}@{}}
\toprule
\textbf{ID} & \textbf{Name} & \textbf{Automation} & \textbf{Primary Concern} \\
\midrule
C01 & Success Integrity & Automated & Success returned after critical failure \\
C02 & Audit / Evidence Integrity & Automated & Audit or evidence loss occurs silently \\
C03 & Broad Exception Suppression & Automated & Exceptions swallowed or neutralized \\
C04 & Distributed Fallback & Automated (heuristic) & Fallback scattered, weakens guarantees \\
C05 & Bypass / Override Paths & Automated & Flags or overrides disable safeguards \\
C06 & Ambiguous Return Contracts & Automated & \texttt{None}/\texttt{null} blurs absence and failure \\
C07 & Parallel Logic Drift & Human review only & Divergent paths produce inconsistent semantics \\
C08 & Unsupervised Background Tasks & Automated & Async work lacks supervision \\
C09 & Environment-Dependent Safety & Automated & Safety relaxed in dev/staging paths \\
C10 & Startup Integrity & Automated & Init catches failure and continues \\
C11 & Deterministic Reasoning Drift & Automated & Decision paths use non-deterministic settings \\
C12 & Source-to-Output Lineage & Human review only & Outputs lack traceable origin \\
C13 & Confidence Misrepresentation & Automated & Degraded output presented without confidence posture \\
C14 & Test Coverage Asymmetry & Automated & Failure-path tests lag happy-path tests \\
C15 & Retry / Idempotency Drift & Automated & Retries on writes without idempotency controls \\
\bottomrule
\end{tabular}
\end{table}

To illustrate the distinction between AI-typical and human-typical failure handling,
consider C01 (Success Integrity):

\begin{lstlisting}[language=Python, caption={AI-typical: failure concealed}]
def process(data):
    try:
        result = persist(data)
        return {"status": "ok"}   # success returned regardless
    except Exception:
        log.warning("persist failed")
        return {"status": "ok"}   # failure concealed
\end{lstlisting}

\begin{lstlisting}[language=Python, caption={Human-typical: failure propagated}]
def process(data):
    result = persist(data)   
    # raises on failure
    if not result:
        raise PersistenceError("persist returned empty")
    return {"status": "ok"}
\end{lstlisting}

The AI version is not wrong on the happy path.
It is \emph{failure-untruthful}: it returns success after a broken guarantee.

\subsection{Check Co-occurrence Patterns}

The 15 checks are not fully independent.
Certain co-occurrence patterns describe characteristic failure profiles:
\begin{itemize}
  \item \checkid{C03} $+$ \checkid{C01} often indicates explicit failure concealment: exceptions are swallowed to preserve a success-signaling return path
  \item \checkid{C04} $+$ \checkid{C09} often indicates environment-shaped degraded assurance: fallback behavior is distributed and relaxed in non-production environments
  \item \checkid{C02} $+$ \checkid{C10} often indicates a system that can start and report readiness despite losing evidence guarantees at initialization
  \item \checkid{C13} is frequently the epistemic surface that makes the rest look normal: confidence misrepresentation makes other failures invisible to downstream consumers
\end{itemize}

Reviewers should treat co-occurring check failures as failure profiles, not independent findings.

\subsection{Reviewer Instructions}

For each check, reviewers mark:
\begin{itemize}
  \item \texttt{PASS} --- requirement is fully satisfied
  \item \texttt{FAIL} --- requirement is violated (include: file path, line number, description)
  \item \texttt{UNKNOWN} --- cannot be determined without runtime or additional context
\end{itemize}

\textbf{On UNKNOWN:} \texttt{UNKNOWN} does not mean safe.
It means the scanner cannot responsibly automate the conclusion. For governance-critical checks, \texttt{UNKNOWN} should be treated as a conditional \texttt{FAIL} pending manual runtime verification.

% ─────────────────────────────────────────────────────────────────────────────
\section{Empirical Validation}
\label{sec:empirical}
% ─────────────────────────────────────────────────────────────────────────────

This section reports results across three studies.
Study~1 is a deterministic \aira{} audit of a six-system enterprise AI-assisted development environment, presented as an operational characterization of the environment in which the hypothesis emerged.
Study~2 is a rebuilt balanced corpus pilot across AI-attributed and human-authored codebases, designed to establish whether the pattern generalizes beyond the originating environment.
Study~3 is a stricter matched-control replication that rebuilds the human-control arm at larger scale and tests whether the Study~2 signal survives stronger sampling and repo-composition controls.
Together they move the Reward-Shaped Failure Hypothesis from structural argument to replicated measured finding.

The studies differ in unit of scope and should not be compared by file count alone.
Study~1 is a deep environment audit of several related enterprise systems. Study~2 and Study~3 are cross-repository matched comparisons.
Study~3's final strict subset contains 1,910 scanned files across 135 AI-attributed repositories and 279 human-control repositories, making it the broader external-validity test even though Study~1 scans a larger number of files within a bounded development environment.

\subsection{Study Design}

All three studies use the \aira{} deterministic static engine exclusively. LLM-assisted scanning is reported separately in Section~\ref{sec:llm}.
All scans are parser-backed and reproducible. No raw source code, file paths, or snippets are transmitted in research mode.

\textbf{Study~1 --- Enterprise AI-Assisted Environment Audit.} A deterministic scan of six enterprise-grade systems in active AI-assisted development.
The systems include governance, reasoning, synchronization, transcription, and software-engineering support surfaces.
One additional candidate system was excluded before scanning because many known errors had already been repaired, which would have made it a remediated-system sample rather than a clean part of the observed development environment.
System names and identifying details are omitted to avoid conflating the paper's claims with evaluation of specific private implementations.

\textbf{Study~2 --- Corpus Pilot.} A rebuilt balanced comparison of 300 agent-attributed files versus 300 matched human-control files, equally distributed across Python, JavaScript, and TypeScript (100 files per language per arm).
Agent-attributed files were sourced from the AIDev dataset \citep{li2025aidev} by filtering for rows where \texttt{agent\_label} is populated and selecting files from merged PRs.
Human-control files were sourced from repositories whose most recent commit predates January 2022, with no AI tool references in commit messages, README, or repository metadata.
Both arms were stratified to 100 files per language, with file size bounds of 100--2,000 lines to exclude stubs and generated artifacts.
Findings are counted per-file per-check; multiple rule hits within the same file and check are counted individually with no deduplication.
Severity is assigned by the deterministic rule engine heuristic (HIGH: clear failure concealment; MEDIUM: risky ambiguity; LOW: distributed fallback that may be contextually acceptable).
The agent-attributed arm spans 126 repositories; the human-control arm spans 21 repositories --- a structural asymmetry that is discussed as a primary threat to validity in the TypeScript analysis below and in Section~\ref{sec:limitations}.

\textbf{Study~3 --- Strict Matched-Control Replication.} A clean-slate follow-on corpus study using the deduped staged AIDev sample as the AI-attributed source arm and a freshly materialized GitHub PR-file pool as the human-control source arm.
This design was chosen because the available AIDev patch data did not contain usable human PR patch rows.
The human-control extractor collected 1,636 accepted candidate files after AI-attribution exclusion filters and audit logging.
Deterministic matching used language, file-size band, and size-decile cells derived from Arm A, enforced a maximum of four files per repository, and used seed 42. The strict matcher selected 956 human controls from the 1,636-file pool;
one human file was later excluded by the scanner's existing file filter, yielding a fair final comparison of 955 AI-attributed files and 955 human-control files.

\subsection{Study 1: Enterprise AI-Assisted Environment Audit}
\label{sec:study1}

Study~1 formalizes the broader environment that motivated the hypothesis.
The hypothesis did not originate from a single-system post hoc audit.
It emerged from repeated observation across a broader enterprise AI-assisted development environment involving multiple production-grade systems, multiple AI coding tools, and partially overlapping development workflows.
That environment motivated the hypothesis and informed the design of the \aira{} checks.
The formal quantitative evidence reported here is the deterministic audit of the six included systems.
Study~1 should therefore be read as an \emph{operational characterization} of the originating environment rather than as the paper's core comparative proof;
the strongest between-arm test is the public-code matched-control evidence in Study~3.

The six-system scan produced 4,120 findings across 1,643 files: 1,522 \HIGH{}, 1,917 \MED{}, and 681 \LOW{}.
The largest system in the environment remains the anchor case, with 1,189 scanned files and 3,487 findings.
The additional five systems contribute 454 scanned files and 633 findings.
Across the full environment, C07 and C12 returned \texttt{UNKNOWN} as expected because they require human review.

\begin{table}[htbp]
\centering
\caption{Study~1 --- Anonymized Enterprise Environment Audit}
\label{tab:study1-systems}
\small
\begin{tabular}{@{}lrrrrr@{}}
\toprule
\textbf{System} & \textbf{Files} & \textbf{Findings} & \textbf{HIGH} & \textbf{MED} & \textbf{LOW} \\
\midrule
System A (anchor) & 1189 & 3487 & 1332 & 1570 & 585 \\
System B & 192 & 174 & 47 & 97 & 30 \\
System C & 119 & 248 & 74 & 135 & 39 \\
System D & 1 & 11 & 7 & 3 & 1 \\
System E & 78 & 95 & 23 & 57 & 15 \\
System F & 64 & 105 & 39 & 55 & 11 \\
\midrule
\textbf{Total} & \textbf{1643} & \textbf{4120} & \textbf{1522} & \textbf{1917} & \textbf{681} \\
\bottomrule
\end{tabular}
\end{table}

The dominant signals were C03 (Broad Exception Suppression: 1,239 findings), C13 (Confidence Misrepresentation: 715 findings), C04 (Distributed Fallback: 681 findings), and C14 (Test Coverage Asymmetry: 563 findings).
C14 is directly comparable to empirical results on test inclusion in agentic PRs, which show substantial variation in test adoption rates across AI coding agents \citep{haque2026tests}.

A structurally notable property is \emph{severity clustering}: 11 of 13 automated checks produce findings of exactly one severity level.
C01, C02, C05, C09, C10, C11, and C15 are exclusively \HIGH{}. C06, C08, and C13 are exclusively \MED{}.
C04 is exclusively \LOW{}. Only C03 and C14 span multiple levels.
This reflects that different failure modes have consistent consequence profiles --- checks that directly conceal failure produce only \HIGH{} findings;
checks that create ambiguity without direct concealment produce only \MED{}.

Within the anchor system that originally motivated the project, four subsystems produced zero findings: the validation script, database layer, evidence module, and uncertainty module.
This indicates that the framework does not simply flag indiscriminately even on the originating surface.
The pervasiveness of findings across the originating development environment is the key result of Study~1.
The broader environment motivated the hypothesis and informed check design; the formal evidence is the reported deterministic audit.
Study~1 should therefore be read as an operational environment audit, not as a causal AI-vs-human comparison.
That causal and matched question is addressed by Studies~2 and~3.

\smallskip\noindent\textit{Internal validity note.} Study~1 uses a bounded enterprise development environment selected because it is the environment in which the failure pattern was observed, not a randomly sampled population of software projects.
One candidate system was excluded before scanning because known errors had already been repaired.
System D contains only one scanner-eligible file under the current Python/JavaScript/TypeScript rule engine and should not be interpreted as a standalone system-level estimate.

\subsection{Study 2: Rebuilt Balanced Corpus Pilot}
\label{sec:study2}

\begin{table}[htbp]
\centering
\caption{Corpus Pilot --- Arm-Level Overview (300 files per arm)}
\label{tab:corpus-arm}
\small
\begin{tabular}{@{}llrrrr@{}}
\toprule
\textbf{Arm} & \textbf{Source} & \textbf{Files} & \textbf{HIGH} & \textbf{MED} & \textbf{HIGH/file} \\
\midrule
A --- AI-attributed & AIDev (agentic PRs) & 300 & 80 & 148 & 0.267 \\
B --- Human baseline & Pre-2022 repos & 300 & 61 & 195 & 0.203 \\
\bottomrule
\end{tabular}
\end{table}

Under file-weighted analysis (Table~\ref{tab:corpus-arm}), the agent-attributed arm produced 80 \HIGH{}, 148 \MED{}, and 24 \LOW{} findings, compared with 61 \HIGH{}, 195 \MED{}, and 23 \LOW{} in the human-control arm.
This corresponds to 0.267 high-severity findings per file in the agent-attributed arm versus 0.203 in the human-control arm --- a 1.32$\times$ overall excess of high-severity findings in AI-attributed code.

\begin{table}[htbp]
\centering
\caption{High-Severity Findings per File by Language}
\label{tab:corpus-lang}
\small
\begin{tabular}{@{}lrrrrl@{}}
\toprule
\textbf{Language} & \textbf{AI Files} & \textbf{AI HIGH/file} & \textbf{Human Files} & \textbf{Human HIGH/file} & \textbf{Direction} \\
\midrule
JavaScript & 100 & 0.54 & 100 & 0.17 & AI $>$ Human (3.18$\times$) \\
Python & 100 & 0.11 & 100 & 0.12 & Parity \\
TypeScript (raw) & 100 & 0.15 & 100 & 0.32 & Human $>$ AI* \\
TypeScript (excl.\ outlier) & 100 & 0.15 & 99 & 0.10 & AI $>$ Human (1.49$\times$)* \\
\bottomrule
\multicolumn{6}{@{}l@{}}{\small * See Section~\ref{sec:typescript-note} for repo-balance analysis.}
\end{tabular}
\end{table}

The separation was not uniform across languages (Table~\ref{tab:corpus-lang}).
JavaScript showed the clearest difference: 0.54 \HIGH{} per file in the agent-attributed arm versus 0.17 in the human-control arm.
Python was approximately at parity (0.11 versus 0.12).

\subsubsection{TypeScript Repo-Balance Analysis}
\label{sec:typescript-note}

TypeScript initially appeared to cut against the hypothesis (0.15 versus 0.32 \HIGH{}/file).
However, follow-up analysis showed this counter-signal is not robust. The human-control TypeScript arm spanned only 6 repositories, and a single file from \texttt{onlook-dev/onlook} contributed 22 of the 32 human-control TypeScript \HIGH{} findings.
Excluding that outlier reduces the human-control TypeScript \HIGH{} rate from 0.32 to 0.101 per file.
A bootstrap sensitivity analysis (5,000 draws of 6 agent repositories from the 55 available) found $P = 0.51$ for the agent mean equaling or exceeding the outlier-excluded human mean.
The apparent TypeScript reversal is not stable under repo-balanced analysis.

\begin{table}[htbp]
\centering
\caption{Check-Level Failure Counts by Arm (300 files per arm)}
\label{tab:corpus-checks}
\small
\begin{tabular}{@{}lrrl@{}}
\toprule
\textbf{Check} & \textbf{Arm A (AI)} & \textbf{Arm B (Human)} & \textbf{Direction} \\
\midrule
exception\_handling (C03) & 61 & 43 & AI $>$ Human \\
confidence\_representation (C13) & 11 & 7 & AI $>$ Human \\
fallback\_control (C04) & 14 & 12 & AI $>$ Human \\
\midrule
background\_tasks (C08) & 15 & 28 & Human $>$ AI \\
environment\_safety (C09) & 6 & 8 & Human $>$ AI \\
return\_contracts (C06) & 7 & 9 & Human $>$ AI \\
\bottomrule
\end{tabular}
\end{table}

At the check level (Table~\ref{tab:corpus-checks}), the strongest recurring support for the hypothesis concentrated in exception-handling-related patterns.
The agent-attributed arm exceeded the human-control arm on \checkid{exception\_handling} (C03: 61 versus 43), \checkid{confidence\_representation} (C13: 11 versus 7), and \checkid{fallback\_control} (C04: 14 versus 12).
By contrast, the human-control arm exceeded the agent-attributed arm on \checkid{background\_tasks} (28 versus 15), \checkid{environment\_safety} (8 versus 6), and \checkid{return\_contracts} (9 versus 7).

\smallskip\noindent\textit{Internal validity note.} Study~2 is file-balanced but not repo-balanced (126 agent repositories vs.\ 21 human repositories).
Language-level comparisons are therefore more reliable for JavaScript, where the arm composition is closer to symmetric, than for TypeScript, where control-side concentration effects are material.
The check-level counts should be treated as directional rather than precise.

\subsection{Study 3: Strict Matched-Control Replication}
\label{sec:study3}

Study~3 addresses the main threat exposed by Study~2: the human-control arm was harder to construct and more composition-sensitive than the AI-attributed arm.
The follow-on design therefore rebuilt Arm B from scratch rather than extending the 600-file pilot.
The final candidate pool contained 1,636 accepted human-control files across JavaScript, Python, and TypeScript, with zero remaining strict matching-cell gaps at the extraction stage.
Matching then enforced a four-file repository cap and selected 956 controls.
After scanner filtering, the fair final analysis subset contains 955 AI-attributed files and 955 human-control files.

\begin{table}[htbp]
\centering
\caption{Strict Matched-Control Replication --- Arm-Level Overview (955 files per arm)}
\label{tab:study3-arm}
\small
\begin{tabular}{@{}lrrrrr@{}}
\toprule
\textbf{Arm} & \textbf{Files} & \textbf{HIGH} & \textbf{MED} & \textbf{LOW} & \textbf{HIGH/file} \\
\midrule
A --- AI-attributed & 955 & 415 & 716 & 65 & 0.435 \\
B --- Human control & 955 & 231 & 617 & 66 & 0.242 \\
\bottomrule
\end{tabular}
\end{table}

Table~\ref{tab:study3-arm} reports the final strict subset.
The AI-attributed arm produced 415 \HIGH{} findings, compared with 231 in the matched human-control arm.
This corresponds to 0.435 high-severity findings per file versus 0.242, or a 1.80$\times$ excess in the AI-attributed arm.

\begin{table}[htbp]
\centering
\caption{Strict Matched-Control Replication --- High-Severity Findings per File by Language}
\label{tab:study3-lang}
\small
\begin{tabular}{@{}lrrrrl@{}}
\toprule
\textbf{Language} & \textbf{AI Files} & \textbf{AI HIGH/file} & \textbf{Human Files} & \textbf{Human HIGH/file} & \textbf{Direction} \\
\midrule
JavaScript & 327 & 0.682 & 327 & 0.367 & AI $>$ Human (1.86$\times$) \\
Python & 311 & 0.293 & 311 & 0.129 & AI $>$ Human (2.27$\times$) \\
TypeScript & 317 & 0.319 & 317 & 0.224 & AI $>$ Human (1.42$\times$) \\
\bottomrule
\end{tabular}
\end{table}

The language-level results are important because they resolve the main ambiguity in Study~2.
In the pilot, JavaScript carried the strongest signal, Python was near parity, and TypeScript was unstable under repo-balanced sensitivity analysis.
In Study~3, all three languages point in the same direction (Table~\ref{tab:study3-lang}).
The effect is largest in Python by ratio, strongest in absolute rate in JavaScript, and still present in TypeScript.

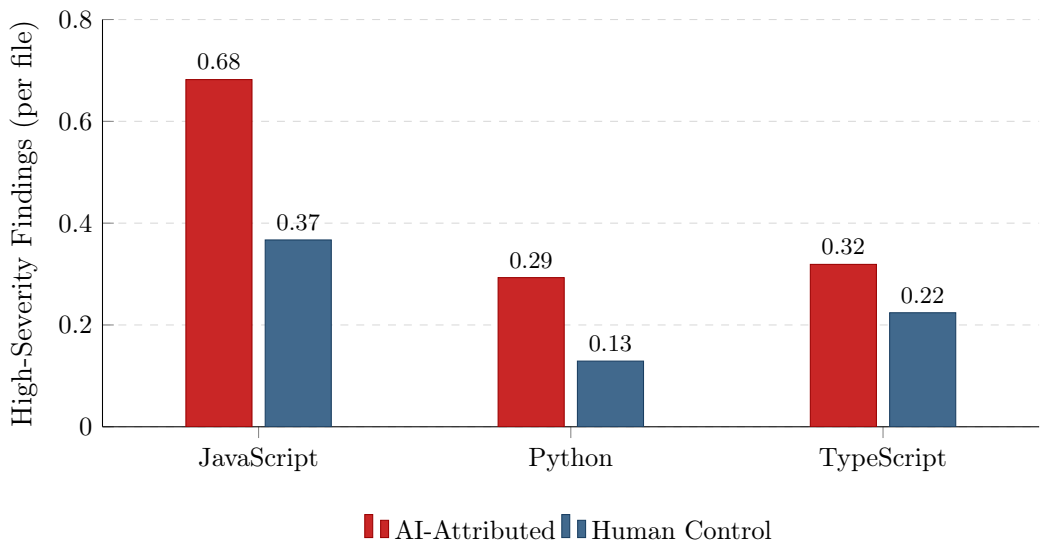
\begin{figure}[htbp]
\centering
\begin{tikzpicture}
\begin{axis}[
    ybar=5pt, 
    bar width=25pt,
    width=0.85\textwidth,
    height=7cm,
    enlarge x limits=0.25,
    legend style={
        at={(0.5,-0.2)}, 
        anchor=north, 
        legend columns=-1, 
        draw=none, 
        fill=none,
        font=\small
    },
    ylabel={High-Severity Findings (per file)},
    symbolic x coords={JavaScript, Python, TypeScript},
    xtick=data,
    nodes near coords,
    nodes near coords align={vertical},
    nodes near coords style={font=\footnotesize},
    ymin=0, ymax=0.8, 
    axis lines*=left,
    ymajorgrids=true,
    grid style={dashed, gray!30},
    tick label style={font=\small}
]

% AI-Attributed Data (Red)
\addplot[fill=aira-red!85, draw=aira-red!80!black] coordinates {
    (JavaScript,0.682) 
    (Python,0.293) 
    (TypeScript,0.319)
};

% Human Control Data (Blue)
\addplot[fill=aira-blue!85, draw=aira-blue!80!black] coordinates {
    (JavaScript,0.367) 
    (Python,0.129) 
    (TypeScript,0.224)
};

\legend{AI-Attributed, Human Control}
\end{axis}
\end{tikzpicture}
\caption{Study 3: High-Severity Findings per File by Language (Strict Matched-Control Replication). AI-attributed code shows a directional excess of high-severity fail-soft patterns across all three evaluated languages.}
\label{fig:study3-bar}
\end{figure}

\begin{table}[htbp]
\centering
\caption{Strict Matched-Control Replication --- Selected Check-Level Failure Counts}
\label{tab:study3-checks}
\small
\begin{tabular}{@{}lrrl@{}}
\toprule
\textbf{Check} & \textbf{Arm A (AI)} & \textbf{Arm B (Human)} & \textbf{Direction} \\
\midrule
exception\_handling (C03) & 263 & 185 & AI $>$ Human \\
environment\_safety (C09) & 23 & 13 & AI $>$ Human \\
background\_tasks (C08) & 80 & 96 & Human $>$ AI \\
confidence\_representation (C13) & 43 & 45 & Approx.\ parity \\
fallback\_control (C04) & 33 & 35 & Approx.\ parity \\
return\_contracts (C06) & 30 & 32 & Approx.\ parity \\
\bottomrule
\end{tabular}
\end{table}

The check-level pattern remains directional rather than uniform (Table~\ref{tab:study3-checks}).
The largest replicated gap is in exception handling, where AI-attributed files produced 263 failures versus 185 in human controls.
Human controls still exceeded AI-attributed files on background-task failures, and several checks were approximately tied.
This mixed profile is methodologically useful: the Study~3 signal is not a scanner-wide inflation artifact, but a concentrated difference in the failure modes most closely related to fail-soft behavior.

\smallskip\noindent\textit{Internal validity note.} Study~3 improves the control design but does not eliminate all sampling constraints.
The strict matcher targeted 1,000 controls but selected 956 under the four-file repository cap, and the final scanner comparison uses a 955-per-arm subset after one human-control file was filtered.
The correct interpretation is therefore a strict matched replication at 955 files per arm, not a perfect 1,000/1,000 match.

\subsection{LLM vs.\ Static Comparison}
\label{sec:llm}

To examine whether the failure-suppression pattern extends to the evaluation layer, we ran a parallel hybrid scan using minimax-m2:cloud via Ollama as a secondary evaluator across anchor-system surfaces from the Study~1 environment.
This comparison is presented as a secondary exploratory finding, not as primary evidence for the hypothesis.
It predates the six-system Study~1 expansion, so its static counts refer to the earlier anchor-system scan rather than the full Study~1 environment aggregate.
The LLM evaluator was prompted with a normalized JSON audit contract requesting structured check-level output;
C07 and C12 were forcibly kept \texttt{UNKNOWN} in all LLM output to prevent the evaluator from automating human-review-only conclusions.
Results at repo scale were subject to prompt truncation; single-file results were not truncated and are the most interpretable.

\begin{table}[htbp]
\centering
\caption{Anchor-System Static vs.\ LLM Finding Counts Across Governance System Surfaces}
\label{tab:llm}
\small
\begin{tabular}{@{}lrrr@{}}
\toprule
\textbf{Surface} & \textbf{Files} & \textbf{Static Findings} & \textbf{LLM Findings} \\
\midrule
Anchor repo (truncated) & 1,124 & 3,297 & 0 \\
engine/runtime & 28 & 222 & 5 \\
api & 58 & 250 & 6 \\
critic\_infrastructure.py & 1 & 34 & 6 \\
deliberation.py & 1 & 26 & 0 \\
startup.py & 1 & 29 & 0 \\
\bottomrule
\end{tabular}
\end{table}

\begin{table}[htbp]
\centering
\caption{Check-Level Suppression: Static FAIL vs.\ LLM PASS (Single Files, No Truncation)}
\label{tab:suppression}
\small
\begin{tabular}{@{}lrrrr@{}}
\toprule
\textbf{File} & \textbf{Static FAILs} & \textbf{LLM Suppressed} & \textbf{Both FAIL} & \textbf{Suppression Rate} \\
\midrule
critic\_infrastructure.py & 7 & 2 & 5 & 29\% \\
deliberation.py & 7 & 7 & 0 & 100\% \\
startup.py & 8 & 8 & 0 & 100\% \\
\bottomrule
\end{tabular}
\end{table}

The single-file comparisons (Tables~\ref{tab:llm} and~\ref{tab:suppression}) are the most controlled: no truncation, identical files, identical checks.
For \texttt{deliberation.py}, the static scanner returned \texttt{FAIL} on 7 checks; the LLM returned \texttt{PASS} on all 7 --- 100\% suppression.
For \texttt{startup.py}, static \texttt{FAIL} on 8 checks; LLM \texttt{PASS} on all 8 --- again 100\% suppression.
The suppression is not random across check types. The LLM consistently passed on C02, C03, C04, C05, and C13 --- precisely the checks most directly related to failure concealment.
This result is not primarily a critique of the specific model used.
It demonstrates that the suppression tendency described in the Reward-Shaped Failure Hypothesis appears not only in code generation but in AI-assisted code \emph{evaluation}.
This motivates \aira{}'s deterministic-first design: the audit instrument must not be susceptible to the failure mode it is designed to detect.

\subsection{Discussion}

Study~1 establishes that the predicted failure class is real and measurable in the enterprise AI-assisted development environment that motivated the hypothesis.
The severity clustering result --- 11 of 13 automated checks producing single-severity findings --- suggests stable consequence profiles across failure modes.

Study~2 sharpens rather than simplifies the hypothesis. The rebuilt balanced pilot supports a real but conditional signal: AI-attributed code produces modestly more high-severity findings overall (1.32$\times$), with the strongest separation in JavaScript (3.18$\times$) and the clearest check-level support in exception-handling-related patterns.
Python is near parity. The apparent TypeScript reversal is substantially explained by control-side hotspot concentration and does not survive repo-balanced sensitivity analysis.

Study~3 tests whether that conditional signal survives a stronger control-arm design. It does.
Under a stricter matched-control design, AI-attributed files show a 1.80$\times$ high-severity differential over human controls, and all three languages point in the same direction.
The most important change from Study~2 is not merely scale;
it is that the stronger human-control construction removes the TypeScript ambiguity rather than erasing the overall signal.

The data therefore does not support the strongest possible version of the Reward-Shaped Failure Hypothesis --- that AI-attributed code is uniformly hotter across every check and surface.
It does support a more precise and now replicated version: AI-attributed code shows higher rates of high-severity fail-soft patterns under matched comparison, with especially clear support in exception-handling-related behavior.
This conditional account is scientifically stronger than a uniform claim, because it generates testable predictions about where future studies should find the signal and where they should not.

% ─────────────────────────────────────────────────────────────────────────────
\section{Output Specification}
\label{sec:output}
% ─────────────────────────────────────────────────────────────────────────────

All \aira{} audits produce output conforming to the following YAML structure (version~1.2):

\begin{lstlisting}[language=,caption={AIRA v1.2 Output Schema}]
ai_failure_audit:
  audit_version: "1.2"
  success_integrity:        PASS | FAIL | UNKNOWN
  audit_integrity:          PASS | FAIL | UNKNOWN
  exception_handling:       PASS | FAIL | UNKNOWN
  fallback_control:         PASS | FAIL | UNKNOWN
  bypass_controls:          PASS | FAIL | UNKNOWN
  return_contracts:         PASS | FAIL | UNKNOWN
  logic_consistency:        PASS | FAIL | UNKNOWN
  background_tasks:         PASS | FAIL | UNKNOWN
  environment_safety:       PASS | FAIL | UNKNOWN
  startup_integrity:        PASS | FAIL | UNKNOWN
  determinism:              PASS | FAIL | UNKNOWN
  lineage:                  PASS | FAIL | UNKNOWN
  confidence_opacity:       PASS | FAIL | UNKNOWN
  test_coverage_symmetry:   PASS | FAIL | UNKNOWN
  idempotency_safety:       PASS | FAIL | UNKNOWN
  findings:
    - issue: "<description>"
      file: "<path>"
      line: <number>
      severity: HIGH | MEDIUM | LOW
      check: "<check_id>"
\end{lstlisting}

\noindent\textbf{Data Availability.} The \aira{} scanner is released as open source at
\url{https://aira.bageltech.net} and \url{https://github.com/BDB-Labs/aira-scanner}.
Aggregate-only research submissions are supported via the CLI. Raw source code and file paths are never transmitted in research mode.
Benchmark datasets and additional comparative scans are welcomed from the community.

% ─────────────────────────────────────────────────────────────────────────────
\section{Limitations}
\label{sec:limitations}
% ─────────────────────────────────────────────────────────────────────────────

\aira{} should be understood as a research scanner and evolving inspection framework.
The following limitations apply:

\begin{itemize}
  \item Cross-file and repo-level semantic reasoning remains weaker than single-file structural detection.
        Findings are primarily per-file and do not capture emergent failure patterns across module boundaries.
  \item The framework cannot determine that AI wrote the flagged code.
        The hypothesis predicts a higher rate of these patterns in AI-assisted codebases;
        \aira{} measures the patterns, not the authorship.
  \item A \texttt{PASS} result does not mean a system is safe.
        Absence of findings means the targeted patterns were not detected by the current rules --- not that the system exhibits full failure truthfulness.
  \item LLM-assisted scans at repo scale can become optimistic or lossy when prompts are truncated.
        The deterministic engine is the appropriate baseline for research use.
  \item Severity ratings are heuristic.
        Not every flagged pattern is a defect in context.
        Human review of findings is always required before remediation.
  \item Study~1 is an environment audit, not a randomly sampled population estimate.
        The included systems are the enterprise AI-assisted development environment in which the hypothesis emerged.
        System names are withheld, and one remediated candidate system was excluded before scanning.
        Study~1 is therefore best interpreted as an operational characterization of the originating environment rather than as the paper's core comparative proof.
  \item Study~2 reports a rebuilt balanced pilot of 600 files.
        The pilot is file-balanced but not repo-balanced: the agent-attributed arm spans 126 repositories versus 21 for the human-control arm.
        The TypeScript comparison is particularly sensitive to control-side hotspot concentration.
        Study~3 was designed to address this weakness, but the pilot should still be interpreted as a directional precursor rather than the final corpus estimate.
  \item Study~3 improves control construction but remains observational. The final strict comparison is 955 files per arm, not the originally targeted 1,000 files per arm, because a four-file repository cap and scanner filtering reduced the analyzable matched subset.
        The human-control arm is matched on language, file-size band, and size decile, but not on every possible domain, framework, or system-role variable.
  \item C07 and C12 remain human-review-only by design. Automated detection of parallel logic drift and source-to-output lineage failure requires repo-scale semantic comparison not yet implemented.
\end{itemize}

% ─────────────────────────────────────────────────────────────────────────────
\section{Conclusion and Future Work}
\label{sec:conclusion}
% ─────────────────────────────────────────────────────────────────────────────

This paper has introduced the Reward-Shaped Failure Hypothesis --- the proposal that AI coding systems may produce fail-soft code in part because of pressures in their optimization environment --- and \aira{}, a 15-check inspection framework designed to detect the specific failure class this hypothesis predicts.
The central reframing is this: traditional software validation asks whether systems work.
\aira{} asks whether systems tell the truth when they fail. Failure truthfulness is a distinct and auditable system property.
The 15 checks are organized around its measurement.

Study~1 results --- 4,120 findings across 1,643 files in a six-system enterprise AI-assisted development environment, plus an anchor-surface LLM comparison under-reporting at a 44:1 ratio --- provide evidence that the predicted failure class is real within the originating environment and may extend to the evaluation layer.
Study~2, a rebuilt balanced 600-file corpus pilot, extends these findings beyond the originating environment and identifies the control-arm construction problem that a stronger replication must solve.
Study~3 performs that stricter replication: in a 955-file-per-arm matched comparison, AI-attributed code shows 0.435 high-severity findings per file versus 0.242 in human controls, a 1.80$\times$ differential, with the same direction in JavaScript, Python, and TypeScript.

The hypothesis is therefore supported as a real but conditional pattern --- not a uniform field-wide law, but a measurable and reproducible tendency whose strength varies by language, check type, and code surface.
The strongest replicated evidence is not that AI-attributed code fails every check more often;
it is that AI-attributed code shows higher rates of high-severity fail-soft patterns under increasingly strict matched comparison.

The three most important next steps are: (1) domain and system-type stratification within the matched public-code corpus;
(2) a controlled base-vs-RLHF model comparison on structured generation tasks, to move the hypothesis from observational evidence to direct experimental proof;
and (3) expansion of the deterministic rule engine to support cross-file and repo-level reasoning.
\aira{} is available as an open-source tool at \url{https://aira.bageltech.net} and \url{https://github.com/BDB-Labs/aira-scanner}. The research collection pipeline supports aggregate-only community submissions.
Contributions are welcomed, especially comparative scan datasets, rule extensions, and calibration studies.

% ─────────────────────────────────────────────────────────────────────────────
\section*{Acknowledgments}
% ─────────────────────────────────────────────────────────────────────────────

\aira{} was developed through direct observation during construction of Constitutional AI governance frameworks with heavy AI coding assistance.
The author thanks the broader AI safety and software reliability research communities whose parallel work motivates the bridge this paper attempts to build.

% ─────────────────────────────────────────────────────────────────────────────
\bibliographystyle{plainnat}

\end{document}